\pdfoutput=1
 \documentclass[aps, prapplied,twocolumn,showpacs,superscriptaddress,preprintnumbers, longbibliography,nofootinbib]{revtex4-2}
\usepackage[normalem]{ulem}
\usepackage{amsmath}
\usepackage{amssymb}
\usepackage{epsfig}

\usepackage{url} 

\usepackage{graphicx}
\usepackage{hyperref}
\usepackage{tabu}
\usepackage{boldline}
\usepackage{xspace}
\usepackage{slashed}
\usepackage{multirow}
\usepackage{booktabs}
\usepackage{diagbox}
\usepackage{tabularx}
\usepackage{comment}
\usepackage{floatrow}
\newfloatcommand{capbtabbox}{table}[][\FBwidth]
\usepackage{color}
\usepackage[normal]{subfigure}
\usepackage[table]{xcolor}
\usepackage{enumitem}
\usepackage[utf8]{inputenc}
\usepackage{colortbl}
\definecolor{nicered}{rgb}{0.7,0.1,0.1}
\definecolor{nicegreen}{rgb}{0.1,0.5,0.1}
\definecolor{red}{rgb}{1.0, 0, 0}
\definecolor{pink}{RGB}{255, 0, 145}

\usepackage{lineno}
\definecolor{LightCyan}{rgb}{0.88,1,1}
\definecolor{piggypink}{rgb}{0.99, 0.87, 0.9}
\definecolor{applegreen}{rgb}{0.55, 0.71, 0.0}
\definecolor{darkpastelgreen}{rgb}{0.01, 0.75, 0.24}
\definecolor{green-yellow}{rgb}{0.68, 1.0, 0.18}

\newcommand{\beq}{\begin{equation}}
\newcommand{\eeq}{\end{equation}}
\newcommand{\beqa}{\begin{eqnarray}}
\newcommand{\eeqa}{\end{eqnarray}}

\usepackage[normalem]{ulem}
\usepackage{amsmath}
\usepackage{amssymb}
\usepackage{epsfig}
\usepackage{graphicx}
\usepackage{hyperref}
\usepackage{tabularx}
\usepackage{tabu}
\usepackage{boldline}
\usepackage{xspace}
\usepackage{slashed}
\usepackage{multirow}
\usepackage{diagbox}
\usepackage{tabularx}
\usepackage{colortbl}
\usepackage{xcolor}

\usepackage{comment}
\graphicspath{{figs/}}

\begin{document}


\title{The Skipper CCD for low-energy threshold particle experiments above ground}

\author{Guillermo Fernandez Moroni}
\affiliation{\normalsize\it Fermi National Accelerator Laboratory, PO Box 500, Batavia IL, 60510, USA.}
\author{Fernando Chierchie}
\affiliation{\normalsize\it Instituto de Inv. en Ing. Eléctrica ``Alfredo Desages'' (IIIE), Dpto. de Ing. Eléctrica y de Computadoras. CONICET and Universidad Nacional del Sur (UNS), Bahía Blanca, Argentina.}
\author{Javier Tiffenberg}
\affiliation{\normalsize\it Fermi National Accelerator Laboratory, PO Box 500, Batavia IL, 60510, USA.}
\author{Ana Botti}
\affiliation{\normalsize\it Department of Physics, FCEN, University of Buenos Aires and IFIBA, CONICET, Buenos Aires, Argentina.}

\author{Mariano Cababie}
\affiliation{\normalsize\it 
Department of Physics, FCEN, University of Buenos Aires and IFIBA, CONICET, Buenos Aires, Argentina.}
\affiliation{\normalsize\it 
Fermi National Accelerator Laboratory, PO Box 500, Batavia IL, 60510, USA.}
\author{Gustavo Cancelo}
\affiliation{\normalsize\it Fermi National Accelerator Laboratory, PO Box 500, Batavia IL, 60510, USA.}
\author{Eliana L. Depaoli}
\affiliation{\normalsize\it CNEA - Gerencia de Área Aplicaciones de la Tecnología Nuclear (GAATN) - Gerencia Química Nuclear y Ciencias de la Salud - Dpto. Metrologia de Radioisotopos - Division Metrologia Cientifica Centro Atómico Ezeiza. }

\author{Juan Estrada}
\affiliation{\normalsize\it Fermi National Accelerator Laboratory, PO Box 500, Batavia IL, 60510, USA.}
\author{Stephen E. Holland}
\affiliation{\normalsize\it 
Lawrence Berkeley National Laboratory, One Cyclotron Road, Berkeley, California 94720, USA.}
\author{Dario Rodrigues}
\affiliation{\normalsize\it 
Department of Physics, FCEN, University of Buenos Aires and IFIBA, CONICET, Buenos Aires, Argentina.}
\affiliation{\normalsize\it 
Fermi National Accelerator Laboratory, PO Box 500, Batavia IL, 60510, USA.}
\author{Iván Sidelnik}
\affiliation{\normalsize\it CONICET y CNEA - Comisión Nacional de Energía Atómica, Departamento de física de neutrones, Centro Atómico Bariloche, Av. Bustillo 9500, Bariloche, Argentina.}
 \author{Miguel Sofo Haro}
\affiliation{\normalsize\it 
Fermi National Accelerator Laboratory, PO Box 500, Batavia IL, 60510, USA.}
\affiliation{\normalsize\it Centro At\'omico Bariloche, CNEA/CONICET/IB, Bariloche, Argentina.}
\author{Leandro Stefanazzi}
\affiliation{\normalsize\it 
Fermi National Accelerator Laboratory, PO Box 500, Batavia IL, 60510, USA.}
\author{Sho Uemura}
\affiliation{\normalsize\it 
 School of Physics and Astronomy, 
 Tel-Aviv University, Tel-Aviv 69978, Israel.}

\date{\today}

\begin{abstract}
We present experimental results using a single-electron resolution Skipper-CCD running above ground level to demonstrate the potential of this technology for its use in reactor neutrino observations and other low-energy particle interaction experiments. Operating conditions and event-selection criteria are provided to decouple most of the background rate at low energies. Our final results for events with energies as low as $5$ ionized electron-hole pairs show that the exponentially increasing rate of events seen in other technologies is not present in our data. This demonstrates that the Skipper CCD proves to be among the best options to measure low energy and weakly interacting particles at ground level.

\end{abstract}

\maketitle


\section{THICK FULLY DEPLETED CCDS FOR PARTICLE DETECTION EXPERIMENTS}\label{sec:intro}

Charged Coupled Devices (CCD) with low readout noise and large active volume have been identified among the most promising detector technologies for the low-mass direct dark-matter search experiments, probing electron and nuclear recoils from sub-GeV mass \cite{Aguilar-Arevalo:2016ndq, Aguilar-Arevalo:2016zop,Aguilar-Arevalo:2019wdi, Crisler:2018gci,Abramoff:2019dfb}. The recent development of the Skipper-CCD \cite{skipper_2012,Tiffenberg:2017aac,Sensei2020}  demonstrated the capability to measure ionization events with sub-electron noise extending the reach of this technology to unprecedented low energies. This enabled world leading results for dark matter searches at underground facilities \cite{Sensei2020}. Experiments based on this technology are planned for the coming years with active masses going from $100$ grams to several kilograms \cite{DAMIC-M,Oscura}, running at underground facilities that allows to rule out most of the cosmic background. 
At the same time, the low-noise CCD technology has been implemented in low-energy neutrino experiments \cite{CONNIE_2019,CONNIE_2020} above surface; implementing Skipper-CCDs is also planned for future developments \cite{Violeta}.

CCD sensors to be used in future instruments are part of a significant R\&D effort \cite{DAMIC-M,Oscura,Violeta}. One of the most important performance requirement to fully exploit the single counting capability of the Skipper-CCD is the control of the background signals below $1$ keV. Rates of single electron depositions larger than expected have been observed in \cite{Sensei2020}, and its possible link to higher-energy particles has been summarized in \cite{du2020sources}. Moreover, recent studies on regions in the CCD with low collection efficiency \cite{pcc_2021} show how large-energy ionization events could produce low-energy signals in the detector. For this reasons, experiments running above ground are more challenging due to the larger rate of atmospheric high-energy particles \cite{Asorey2021,Sarmiento2020}. 
Several new low-threshold technologies observe a rapidly increasing rate of background events towards low energies (below 1 keV). This effect is more evident when detectors are not operated underground. Larger background rates result in limiting the reach of the low-threshold capability. The first edition of the EXCESS workshop in 2021 \cite{excess_workshop_2021} exhibit the effort of the community to overcome this issue.

In \cite{Fernandez-Moroni:2020} it is shown that a good control of the background starting at 15 eV up to 500 eV is a key factor for new physics searches using neutrinos at nuclear reactors. Moreover, other studies such as the silicon energy absorption for photons and neutrons scatterings are critical and require Skipper-CCDs running at the surface level with controlled levels of background. 
It is fairly known that the long readout time of these sensors does not allow the use of active veto systems. For this reason, passive shield configurations together with event selection criteria based on the spatial resolution of thick fully-depleted CCDs are used to reject undesired signals.

In this article we study the use of Skipper-CCDs as a potential low-threshold technology for experiments above surface. We discuss the main sources of background found on the measurements. We propose a detector operation scheme and selection criteria to reject these contributions. 
We demonstrate that with the proposed technique, the low energy excess in the background rate seen by other experiments \cite{excess_workshop_2021} is not present using the Skipper-CCD technology, even down to events producing an ionization of $5$ electron-hole pairs (just electrons from now on). 
Finally, we compare the background measurements observe by other low threshold technologies already published that have participated in the EXCESS workshop.

\section{Description of the experiment}

\begin{figure}[htbp]
\includegraphics[width=0.5\textwidth]{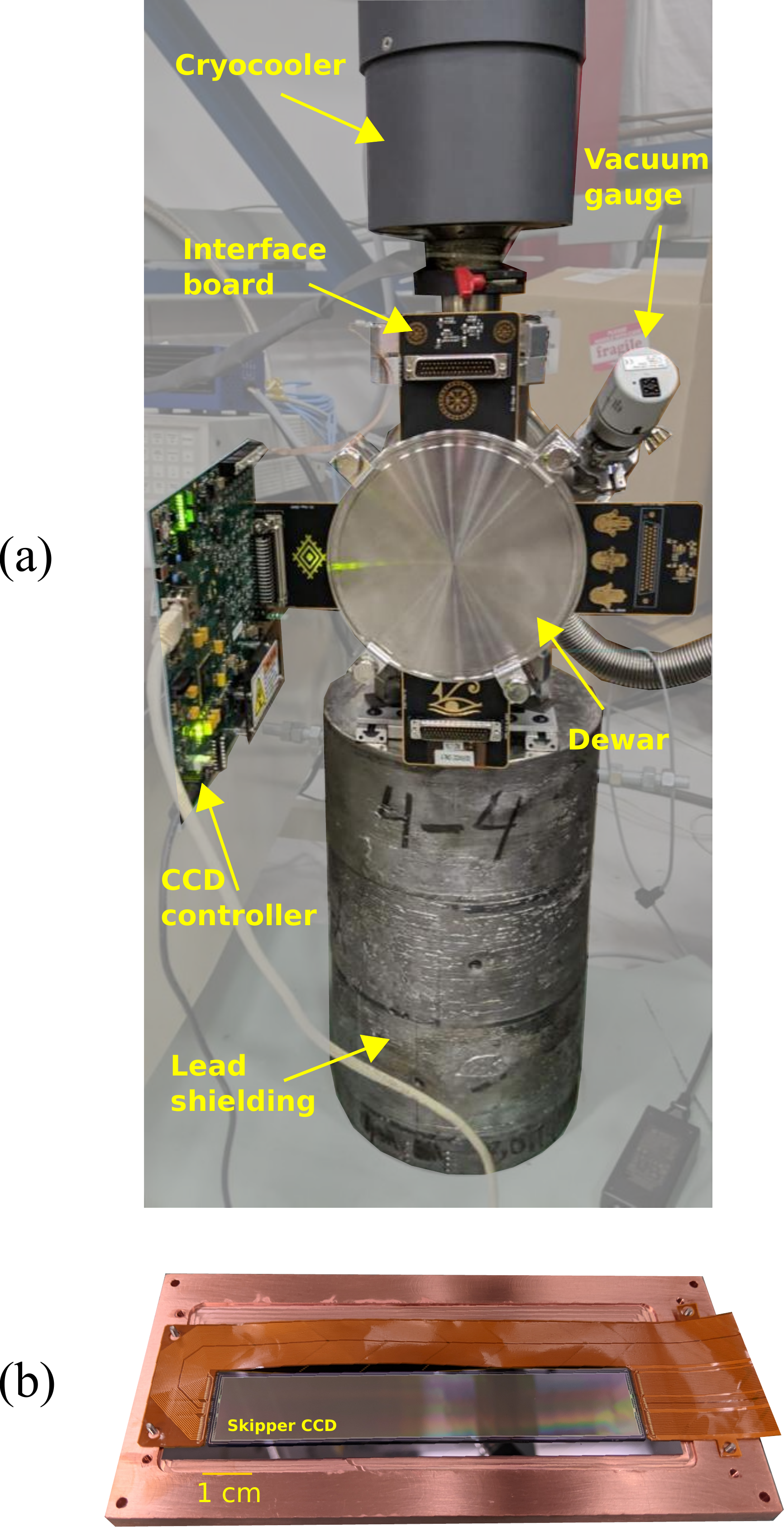}
\caption{(a) Setup used for the experiment with a short description of the main components. (b) A picture of the CCD installed on the Copper tray. An extra Copper plate that covers the top part of the sensor is not presented in the image.}
\label{fig:setup}
\end{figure}

Figure \ref{fig:setup} (a) shows the experimental setup where some of the main components are labeled. One Skipper-CCD (shown in Fig. \ref{fig:setup} (b)) was operated at a temperature of $140$ K using a Sunpower cryocooler \cite{Sunpower_2021}. The CCD is glued to a silicon substrate that sits on a Copper tray for mechanical support as well as thermal connectivity. The CCD is placed in an extension of the dewar that fits inside a lead cylinder. A lead cap on top of the sensor (inside the dewar extension) completes the lead shield of two inches of thickness around the device. There was no radio purity selection of materials inside the shield. The CCD was designed by the LBNL Microsystems Laboratory (MSL)~\cite{LBNL-MSL} and fabricated at Teledyne-DALSA. It has $6144$ columns by $1024$ rows with pixels of $15$ $\mu$m by $15$ $\mu$m with a thickness of $675$ $\mu$m. It is read by four amplifiers, one on each corner, using a Low Threshold Acquisition (LTA) controller \cite{cancelo2021low}. Two quadrants presented larger readout noise and were not used for the analysis in the following sections. The sensor was operated at sub-electron noise by averaging $300$ measurements of the charge in each pixel \cite{Tiffenberg:2017aac} and with a horizontal binning \cite{janesick2001scientific} of $10$ columns.

Two different kind of measurements were performed:
\begin{itemize}
    \item DATA-SET A: Interactions in the active region. $3.21$ days of data were collected from the active region of the sensor in continuous readout mode \cite{Sensei2020}. 
    %
    \item DATA-SET B: Interactions collected in the serial register (also called horizontal register). It consists of charge packets collected only by the serial register, not coming from the active region, from here on called ``serial-register events''\footnote{Later it will be shown that these depositions dominate the low energy spectrum in the output data. This charge is generated in the inactive silicon surrounding the active volume on the edges of the die. The charge can diffuse from the undepleted region and eventually reach the collection region of the pixels in the serial register. For more information about the sensor electrostatic see \cite{Holland:2003}.}. 
    To make an isolated measurement of this source of events, the vertical lines of the sensor were left in idle mode while clocking only the control signals from the serial and output stage. In other words, the sensor is read out as in DATA-SET A but without moving the charge from the active area to the serial register. 2.7 days were collected in this mode. 
\end{itemize}

\section{Detector calibration and noise sources}
\label{calibration}

\begin{figure}[htbp]
\includegraphics[width=1.\textwidth]{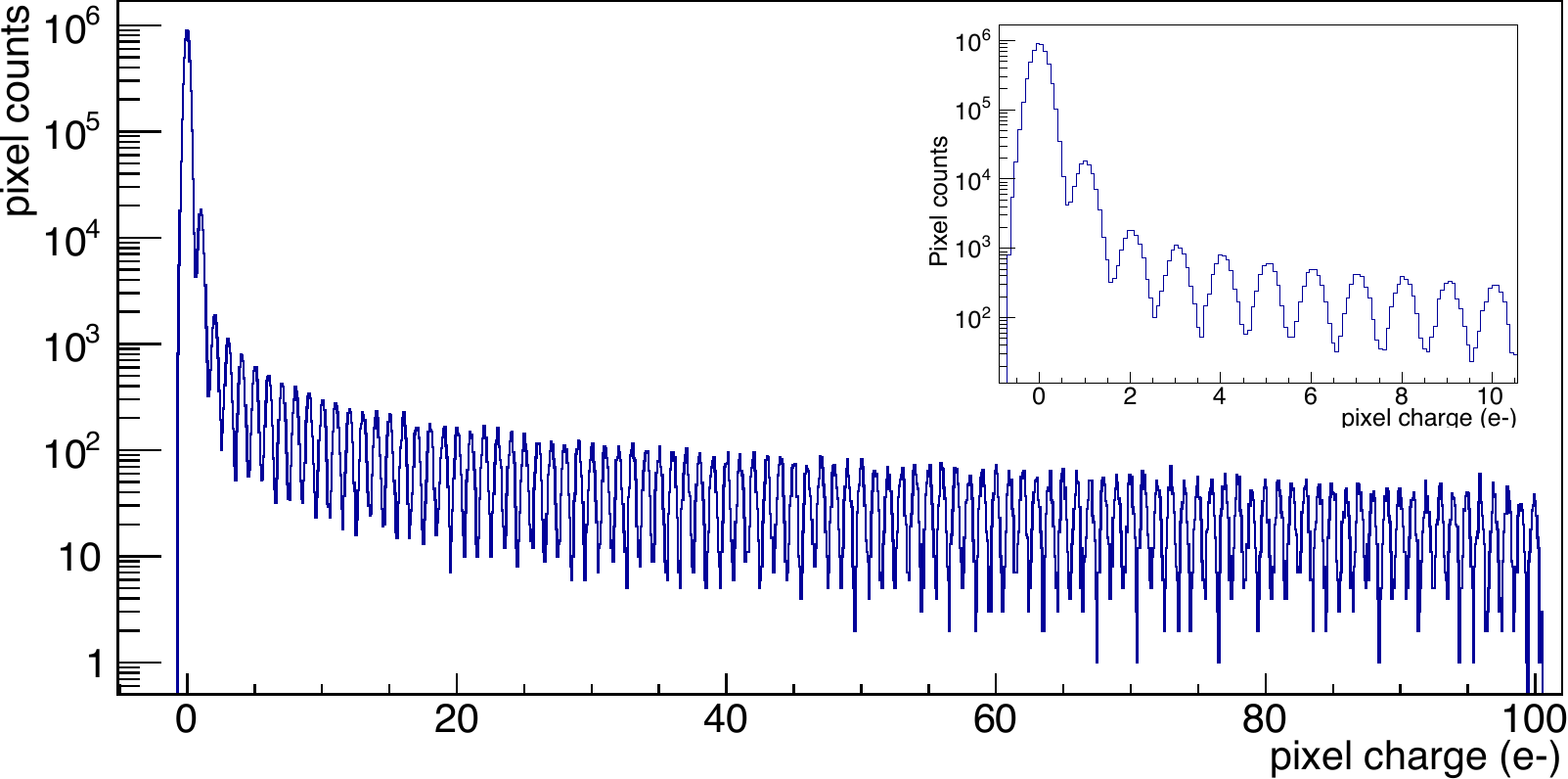}
\caption{Histogram of pixels with charge up to 100$e^-$ after calibration. Single electron discrimination is observed. The inset shows the same histogram for the first ten peaks.}
\label{fig:calibration}
\end{figure}

An absolute energy calibration of the sensor is performed using the electron counting capability. A histogram of the pixel values from the active region (DATA-SET A) is produced as shown in Fig. \ref{fig:calibration}. Each peak correspond to a discretized number of electrons in the pixel after the calibration and is fitted using a Gaussian distribution whose mean value is used to build a look-up table (digital unit vs. electrons) to calibrate the sensor up to around $700$ $e^-$. 
To calibrate the sensor at $2146$ $e^-$, single pixel X-ray events with energy of 8.048 keV produced by the fluorescence of the surrounding copper material are also used.
We assume an average energy deposition per collected electron of 3.75 eVee \cite{RODRIGUES2021}.

The readout noise of the sensor, evaluated as the standard deviation of the values of the empty pixels ($0$ $e^-$ peak in Fig. \ref{fig:calibration}), is 0.165 $e^-$ and 0.167 $e^-$ for the two quadrants in use. The average single electron rate per pixel measured are $0.01$ $e^-$/pix and $0.009$ $e^-$/pix after binning in each quadrant, with a contribution from the exposure dependent charge production \cite{cababie_2021} of $0.018$ $e^-$/pix/day (pixels before binning). The diffusion charge transport of carriers in the bulk of the sensor was measured and it turned out to be similar to the previously published in \cite{Sensei2020}.

\section{Study of the extracted interaction}

\begin{figure}[htbp]
\includegraphics[width=1.\textwidth]{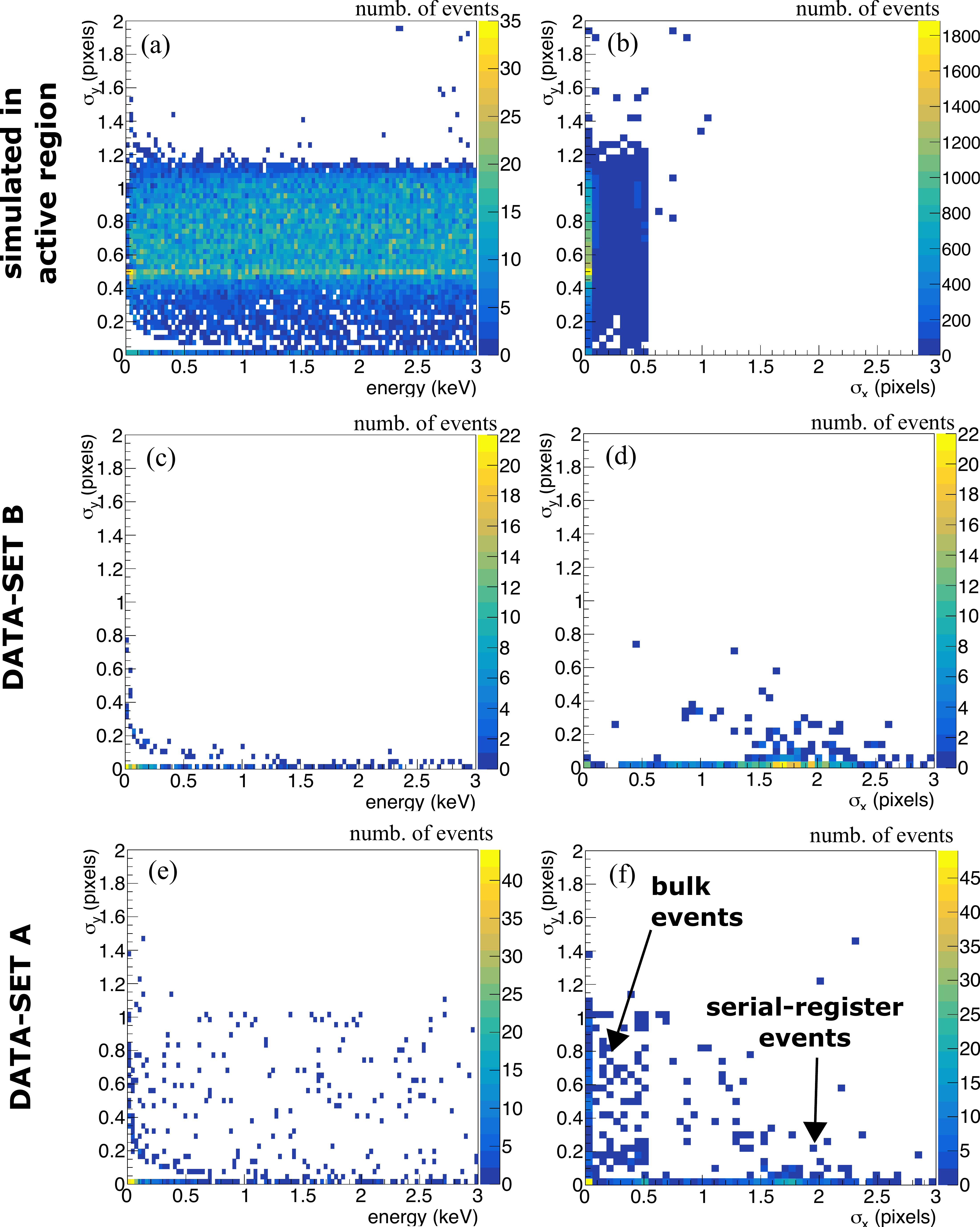}
\caption{Morphology of the events in the output image: (a) and (b) are the expected events from true interactions in the sensor generated from simulations; (c) and (d) serial-register events contributions measured without dumping the collected charge in the active volume to the serial register; and (e) and (f) measured events in the active region of the sensor.}
\label{fig:mophology}
\end{figure}

Figure \ref{fig:mophology} shows the shape and energy distributions of the events analyzed. Left plots show the event size in the $y$-direction (rows of the image), $\sigma_{y}$ (determined as the standard deviation of the charge distribution in the pixels of the event) as a function of energy for three different data sets: the two previously discussed and a third set generated from a simulation. The plots on the right show the relationship between the size ($\sigma_{x}$) in $x$-direction (columns of the image)  and $\sigma_{y}$.

Figure \ref{fig:mophology}(a) and (b) correspond to simulations of particle energy depositions in the active region (bulk) of the sensor. Spatially and energetically uniformly distributed  events ($30000$ events) of charge ionization from $5$ to $800$ electrons were simulated in the bulk \cite{sofo_2020} using the charge transport calibration curve in \cite{Sensei2020}. Since the charge of ten columns is collapsed into one due to the horizontal binning, there is little information left concerning the shape of the event in the $x$-direction of the output images as shown in Fig. \ref{fig:mophology} (b). Some events present $\sigma_{x}>0$. They take place on the edge of the binned regions and have information shared between two binned pixels and their probability of occurrence is low.

From Fig.~\ref{fig:mophology}(a) we can see that the $\sigma_{y}$ distribution is similar for all energies. Most of the events have a $\sigma_{y}$ larger than $0.4$ pixels (see color scale) with a maximum of about $1.1$ pixels due to the maximum expected lateral diffusion. 

Figure \ref{fig:mophology}(c) and (d) shows the distribution of events measured in the serial register (DATA-SET B). A dark-current signal following a Poisson distribution with mean equal to $0.01$ $e^-$/pixel (after binning, as expected in DATA-SET A) was added to emulate both contributions together. In contrast to the events in the bulk, these ones show a $\sigma_{y}$ distribution close to $0$. Since the charge is only collected in one row (the serial register), serial-register events have practically no spatial information in the $y$-direction. They extend on the $x$-direction projecting on the $\sigma_{x}$ axis. 
Low-energy events may appear with an increased size in the $y$-direction due to dark counts in neighboring pixels paired in the same cluster.

The third group in Fig \ref{fig:mophology}(e) and (f) corresponds to measurements in the active region of the sensor (DATA-SET A). The plots show components of the previously mentioned contribution: events from the bulk (as in the simulations) and from the serial register.
The firsts have sizes below $1.1$ pixels in $y$-direction and smaller than $0.5$ pixel in $x$-direction, as expected from the charge transport calibration and event simulations, while the latter populates both plots for small $\sigma_{y}$.

As shown in the previous analysis, the proposed horizontal binning provides the advantage of orthogonalizing the size information of both sources of events, and therefore a more pure selection of bulk events can be made. It also accelerates the total readout time needed to read the pixel charge while staying in the serial register, and thus it reduces the exposure time to this source of spurious events.

\section{Event selection, final spectrum and comparison with other technologies}

\begin{figure}[htbp]
\includegraphics[width=1\textwidth]{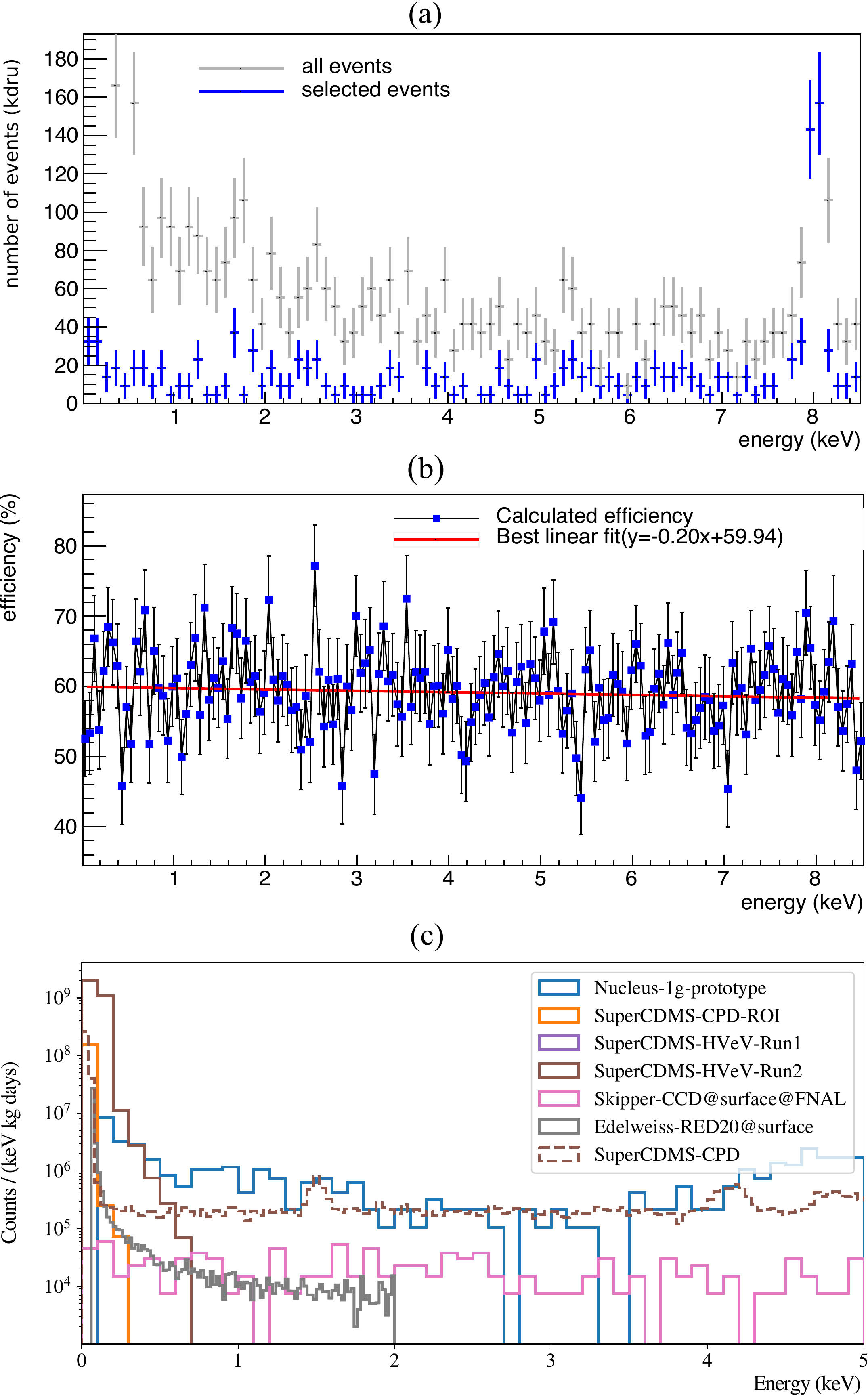}
\caption{(a) Energy spectrum before and after applying the selection criteria to the data (not weighted by the selection efficiency). The first two bins of the spectrum for all the events exceeds the selected maximum value in the $y-axes$. (b) efficiency of the selection cuts. (c) comparison of our results with other low energy technologies presented in \cite{excess_workshop_2021}.}
\label{fig:spectra}
\end{figure}

\begin{figure}[hbp]
\includegraphics[width=1.\textwidth]{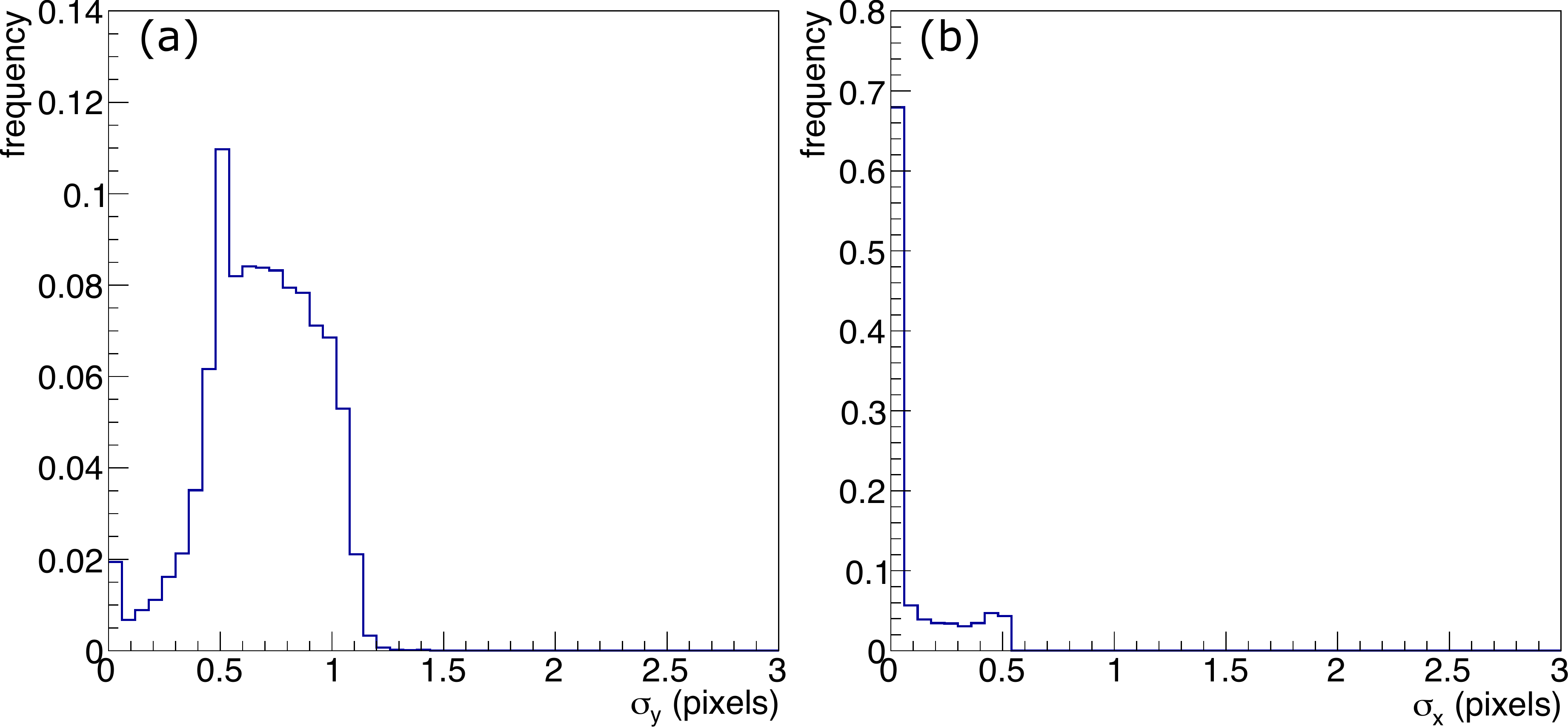}
\caption{Size distribution of events simulated in the active region of the sensor: (a) in the $y$-axis, (b) in the $x$-axis. The bins are normalized by the total number of simulated events.}
\label{fig:size_distribution}
\end{figure}

Figure \ref{fig:spectra}(a) shows the measured differential spectra of events in the active region with $5$ or more electrons of ionized charge (from DATA-SET A). Columns of the CCD that presented high single electron rate (hot columns \cite{janesick2001scientific}) were eliminated from the analysis at an early stage. The remaining active mass of the sensor in use is $0.675$ grams. The gray plot depicts all events reconstructed from the output images showing a monotonic increase of the rate towards low energy. The peak at around $8$ keV corresponds to the fluorescence X-rays from the copper in the mechanical package of the sensor.  
To study the impact of the results obtained in the previous section a selection criteria based on the event shape was performed. Figure \ref{fig:size_distribution} shows the one dimensional distribution of the size, in $x$ and $y$ directions, for the events simulated in the active region of the sensor. Although events in the bulk can show a size in $\sigma_{x}$ up to 0.5 pixels (Fig.~\ref{fig:size_distribution}(a)), the majority of them (72\%) have a width smaller than 0.1 pixels.
A constraint of $\sigma_{x}<0.3$ is used to provide strong rejection to the serial-register and dark-current events. At the same time, only sizes in the y-direction compatible with depositions in the bulk are selected: $0.3<\sigma_{y}<1$. The minimum limit is to prevent counting serial-register events. The spectrum of events after the selection cuts is also shown in blue in Fig.~\ref{fig:spectra}(a). Most of the low energy excess was removed because its contribution was coming from the serial register. The average flux of events with equivalent deposited energy from 5 $e^-$  to 2000 $e^-$ (7.5keV) is $12$ kdru. The plot does not show any significant increase of events towards low energies. The spectrum is normalized by the active mass of the sensor, but not by the efficiency of the selection criteria.


To evaluate the efficiency of the selection criteria, simulated events from interactions spatially and energetically uniformly distributed in the active volume of the sensor were added to the real images. The diffusion transport model of carriers were also used in this simulation. The percentage of reconstructed events passing the selection criteria are shown in Fig. \ref{fig:spectra}(b). Each energy bin is $50$ eV wide and the first bin starts at $15$ eV (equivalent to 5 electrons included). The efficiency is almost flat for all energies showing that the final shape of the spectra is not modified by the selection criteria. The red curve shows a linear fit of the points.

The results in this work compared to other current technologies for low threshold experiment are shown in Fig. \ref{fig:spectra}(c). The plot was produced using the online tool provided by organizers of the EXCESS workshop \cite{excess_workshop_2021}. Only data-sets from measurements above ground with published results are compared in the plot. Table \ref{tab:table1} summarizes the references to the articles were the detector and data analysis information is presented.  All spectra are normalized by efficiency. Our results do not show the rapidly increasing rate of events observed for the other technologies for energies below 1 keV. Moreover, for energies below 500 eV where the neutrino signal from nuclear reactors is expected, our sensor shows the lowest background rate demonstrating the potential of the Skipper-CCD technology for this application and other low-energy measurements.

\begin{table}[hb]
\begin{tabular}{|c|c|}
\hline
Legend in Figure \ref{fig:spectra}(c) & References    \\
\hline
Skipper-CCD@surface@FNAL & our results \\
Nucleus-1g-prototype    & \cite{nucleus_1,nucleus_2,nucleus_3,nucleus_4,nucleus_5}   \\ 
SuperCDMS-CPD-ROI       &  \cite{SuperCDMS-CPD_Fink_2021}   \\
SuperCDMS-HVeV-Run1     & \cite{SuperCDMS-HVeV-Run1}   \\
SuperCDMS-HVeV-Run2     & \cite{SuperCDMS-HVeV-Run2}   \\
Edelweiss-RED20@surface & \cite{Edelweiss-RED20_surface_1,Edelweiss-RED20_surface_2}   \\
SuperCDMS-CPD           & \cite{SuperCDMS-CPD_alkhatib2021light}   \\
\hline
\end{tabular}
\caption{\label{tab:table1} References for the plots spectra shown in Fig. \ref{fig:spectra}(c). References obtained from \cite{excess_workshop_2021}.}.
\end{table}

\section{CONCLUSIONS}

Experimental results using a single-electron resolution Skipper-CCD running above ground have been presented. We demonstrated the potential of this technology for reactor neutrinos and other low-energy particle interaction experiments. The main background source of events has been identified and a technique to decouple its contribution has been presented. Our final result for events with energies of $5$ ionized electron-hole pairs was compared to other state-of-the-art low-energy-threshold technologies showing that the exponentially  increasing rate of events observed is not present in our data as shown in Fig. \ref{fig:spectra}(c).

\begin{acknowledgments}
We thank the SiDet team at Fermilab for the support on the construction and operation of Skipper-CCDs setups, especially Kevin Kuk and Andrew Lathrop.
This work was supported by Fermilab under DOE Contract No.\ DE-AC02-07CH11359. 
This manuscript has been authored by Fermi Research Alliance, LLC under Contract No. DE-AC02-07CH11359 with the U.S.~Department of Energy, Office of Science, Office of High Energy Physics. The CCD development work was supported in part by the Director, Office of Science, of the U.S. Department of Energy under No. DE-AC02-05CH11231. The United States Government retains and the publisher, by accepting the article for publication, acknowledges that the United States Government retains a non-exclusive, paid-up, irrevocable, worldwide license to publish or reproduce the published form of this manuscript, or allow others to do so, for United States Government purposes.
\end{acknowledgments}

\bibliographystyle{apsrev4-1-b.bst}
\bibliography{biblio.bib}

\begin{thebibliography}{38}
\bibitem [{Vio()}]{Violeta}%
  \BibitemOpen
  \href@noop {} {}\bibinfo {note} {The Violeta Collaboration is planning a
  kg-scale Skipper-CCD experiment at a nuclear reactor facility.}\BibitemShut
  {Stop}%
\end{thebibliography}

\end{document}